\documentclass[prb,twocolumn,showpacs]{revtex4-1}
\usepackage{amsmath}
\usepackage{amssymb}
\usepackage{graphicx}
\usepackage{color}
\def\ket#1{\left|#1\right\rangle}
\def\bra#1{\left\langle#1\right|}

\newcommand{\vect}[1]{\boldsymbol{#1}}
\begin{document}

\title{Network Models of Photonic Floquet Topological Insulators}

\author{Michael Pasek}
\email{mpasek@ntu.edu.sg}

\author{Y.~D.~Chong}
\email{yidong@ntu.edu.sg}

\affiliation{School of Physical and Mathematical Sciences and Centre
  for Disruptive Photonic Technologies, Nanyang Technological
  University, Singapore 637371, Singapore}

\begin{abstract}
A recently-proposed class of photonic topological insulators is shown
to map onto Chalker-Coddington-type networks, which were originally
formulated to study disordered quantum Hall systems.  
Such network models are equivalent to the Floquet states of
periodically-driven lattices.  We show that they can exhibit
topologically protected edge states even if all bands have zero Chern
number, which is a characteristic property of Floquet bandstructures.
These edge states can be counted by an adiabatic pumping invariant
based on the winding number of the coefficient of reflection from one
edge of the network.
\end{abstract}


\maketitle

\section{Introduction}
Since the work of Thouless and co-workers \cite{TKNN}, physicists have
recognized that the exotic physics encountered in quantum Hall systems
\cite{MStone}, and more recently topological insulator materials
\cite{Moore, RMPHasan, RMPQi}, is intimately tied to the topological
properties of their bandstructures.  Topological band theory has
since been extended in several interesting directions beyond its
original context.  For example, several groups have shown that when
cold-atom or condensed-matter lattices are subjected to a
time-periodic drive, the resulting Bloch-Floquet states can form
topologically non-trivial bands
\cite{Oka,Inoue,Demler0,Demler,Lindner,Gu}.  These ``Floquet
topological insulators'' \cite{Lindner,Cayssol} exhibit many of the
properties expected of topological materials, such as edge states
which are immune to disorder-induced backscattering, but they also
have some unique and peculiar characteristics of their own; for
example, topologically-protected edge states can exist even when all
the bands have zero Chern number and would thus normally be
considered ``topologically trivial'' \cite{Demler,Levin}.  Topological
bandstructures have also been identified in photonic systems, 
including magneto-optic photonic crystals
\cite{Raghu1,Raghu2,Wang1,Wang2}, cavity QED circuits\cite{LeHur,LeHur1}, metamaterial photonic crystals
\cite{Khanikaev}, and ring resonator lattices
\cite{hafezi,hafezi2,Liang}.  Interest in these systems is driven, in
part, by the possible device applications of topologically-protected
photonic modes (e.g.~the stabilization of slow-light transmission),
and in part by the fundamental interest of combining topological band
physics with optical phenomena (e.g.~gain and nonlinearity).  The
literature on topological photonics has intersected in interesting
ways with the Floquet topological insulator concept: notably, Fang
\textit{et al.}~have studied the Floquet bandstructures formed by
lattices of photonic resonators which are driven periodically
(e.g.,~by electro-optic modulators) \cite{Fan}, while Rechtsman
\textit{et al.}~have experimentally demonstrated a coupled-waveguide
array which acts like a Floquet topological insulator, with adiabatic
wavepacket evolution along a spatially-modulated axis simulating a
time-periodic drive \cite{Szameit}.  We will focus on ring
resonator lattices of the sort studied in
Refs.~\onlinecite{hafezi,hafezi2,Liang}. Such photonic topological insulators
have the technologically desirable properties of being
on-chip, realizable at optical frequencies, and not requiring an
external drive or magnetic field. As originally proposed by
Hafezi \textit{et al.}  \cite{hafezi}, ring resonators are arranged in
a two-dimensional (2D) lattice, and coupled weakly by
specially-engineered waveguides which produce phase shifts
incommensurate with the lattice, analogous to the Landau gauge in the
quantum Hall effect.  Subsequently, it was shown that a topological
bandstructure could be obtained in a lattice with commensurate
couplings \cite{Liang}, analogous to the zero-field quantum
Hall effect \cite{haldane88}.  The transition into the topologically
non-trivial phase occurs by tuning the inter-ring couplings to large
values, such that the system must be treated with transfer matrix
rather than tight-binding methods.  

In this paper, we point out that these resonator-and-waveguide
photonic topological insulators \cite{hafezi,hafezi2,Liang} can be
modeled as networks of the sort developed by Chalker and Coddington in
the 1980s to study the Anderson transition in quantum Hall systems
\cite{ChalkerCo,Kivelson,Lee,Kramer}.  Network models are described by
discrete-time evolution operators in place of Hamiltonians
\cite{klessemet1,HoChalker}, and we show that this allows the Bloch
modes of \textit{periodic} networks to be mapped onto the
Bloch-Floquet states of driven
lattices\cite{klessemet2,janssen,janssenbook}---which, as mentioned
above, have attracted a great deal of recent
attention \cite{Oka,Inoue,Demler0,Demler,Lindner,Gu,Cayssol,Levin}.  To date,
however, ideas from the network model literature have not been
widely employed in the growing Floquet
topological insulator literature.
Furthermore, the network picture allows a topological
invariant to be formulated based on adiabatic pumping \cite{Laughlin,Brouwer}, relating
the number of topologically-protected edge states in the projected
bandstructure to the winding number of a coefficient of reflection
from one edge of the network.

In its original context, a Chalker-Coddington (CC) network model
\cite{ChalkerCo} describes a 2D electron gas subject to a strong
magnetic field and a disorder potential, $V(\vec{r})$, whose
correlation length greatly exceeds the magnetic length. In this
regime, the electron wavefunctions are localized along equipotential
contours of $V(\vec{r})$.  The equipotentials form the directed
links of a network, and each link is associated with an
Aharonov-Bohm phase acquired by the electron amplitude. 
Saddle points of the potential, where the quantum tunneling between adjacent contours (links) can occur, make up the nodes of the network,
which is taken to form a square lattice.
The tunneling between the incoming and outgoing links at each node is
described by a unitary scattering matrix, parameterized by a coupling
strength $\theta$.  One can associate to each network a unitary matrix
relating the inputs and outputs of the entire ensemble of nodes, which
is analogous to a ``discrete-time'' evolution operator
\cite{klessemet1,HoChalker}.  Although the model was originally
formulated for studying the effects of disorder, Ho and Chalker
\cite{HoChalker} subsequently applied the evolution operator analysis
to a periodic square lattice network, and showed that an effective 2D
Dirac Hamiltonian emerges at the critical value $\theta = \pi/4$,
with chiral edge states appearing when $\theta > \pi/4$.  This
result was later rederived, in the context of photonic topological
insulators, in Ref.~\onlinecite{Liang}, together with the bulk and
projected bandstructures.  One of the aims of the present paper is to
clarify the band topology and the nature of the bulk-edge
correspondence in these bandstructures. We will see that the
bandstructures derived in Ref.~\onlinecite{Liang} are characteristic
of ``anomalous Floquet insulators'' (AFI)\cite{Demler,Levin}: all bands have zero Chern
number despite the existence of topologically protected edge states.
We shall also see that network models based on the honeycomb lattice
have richer phase diagrams, containing both ``Chern insulator'' (CI)
phases\cite{haldane88} (where the bands have non-zero Chern number)
and AFI phases. Similar behavior has previously been found in a 2D
hexagonal tight-binding model with periodically-varying hopping
amplitudes\cite{Demler}.

It is interesting to note that in their original context, network
models were intended to be effective descriptions of a system with a
definite underlying Hamiltonian---a non-interacting electron gas in a
magnetic field and disorder potential.  However, the situation is
reversed for photonic resonator lattices: here, the wave amplitude
description of coupled ring resonators \cite{yariv02, yariv_wg} is
valid for arbitrary coupling parameters, and an effective Hamiltonian
(tight-binding) description emerges for weak coupling \cite{hafezi}.

\section{Photonic networks and Floquet maps}
\label{sect:Floquet maps}

\begin{figure}
  \centering\includegraphics[width=0.45\textwidth]{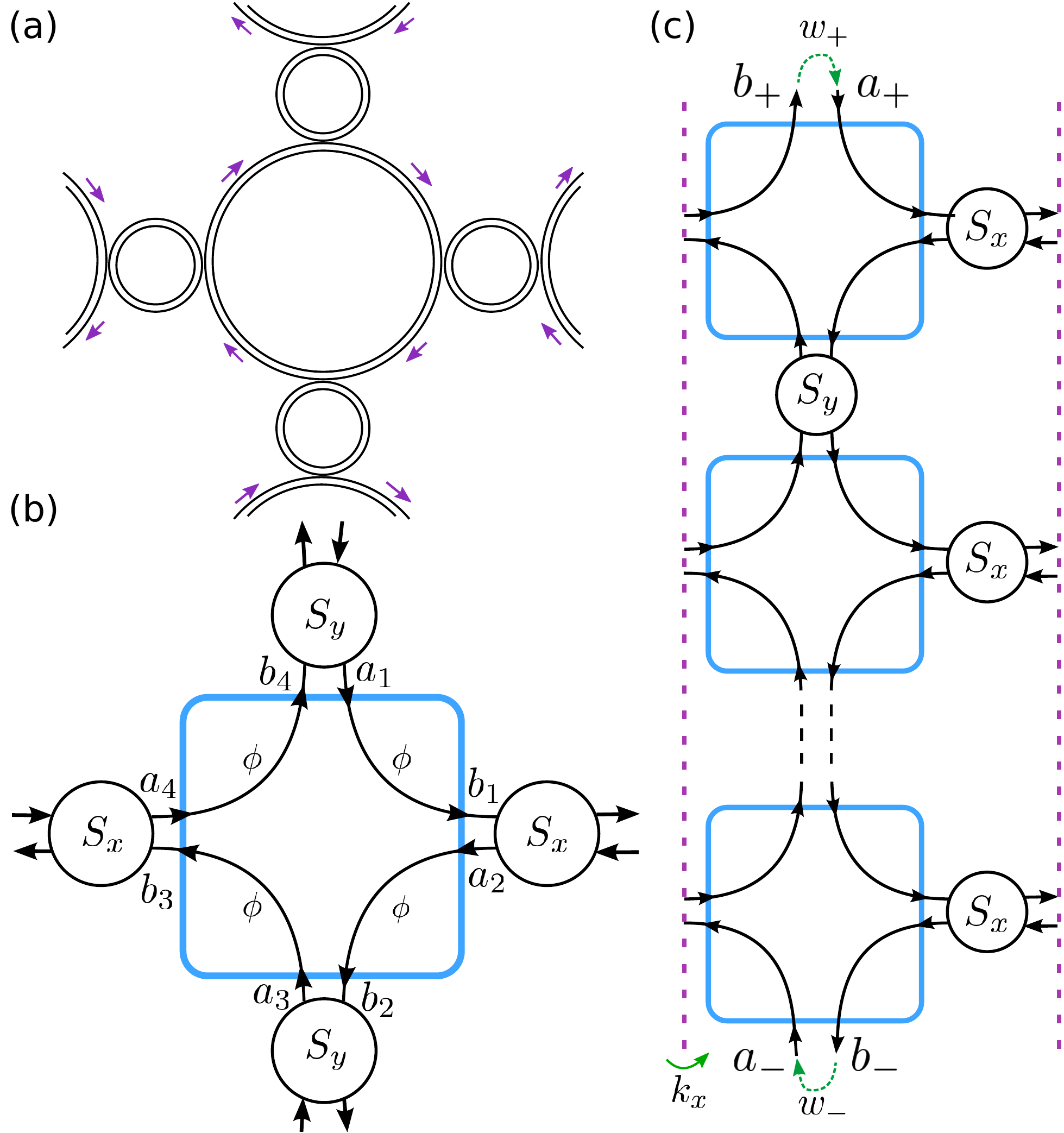}
  \caption{(color online) (a) Schematic of a unit cell in a
    two-dimensional lattice of photonic ring resonators.  (b) The
    equivalent periodic network.  Within the unit cell, we define a
    surface (blue rectangle) which is penetrated by input amplitudes
    $\ket{a}$ and output amplitudes $\ket{b}$, related by $\ket{b} =
    e^{i\phi}\ket{a}$.  These amplitudes also scatter with those of
    neighboring cells, with coupling matrices $S_x$ and $S_y$.  (c) A
    supercell consisting of $N_y$ unit cells joined along the $y$
    direction, with twisted boundary conditions along the $x$
    direction with twist angle $k_x$ and variable phase delays $w_\pm$
    along the upper and lower boundaries.  }
  \label{fig:couplings}
\end{figure}

We begin by examining how a photonic lattice maps onto a network, and
how the network may be described by a unitary evolution matrix.  As
described in Refs.~\onlinecite{hafezi,hafezi2,Liang}, and depicted in
Fig.~\ref{fig:couplings}(a), a photonic topological insulator can be
constructed by a lattice of ring resonators.  Each resonator acts as
an optical waveguide, constraining light to propagating along the
ring.  Each quarter-ring serves as a ``link'' in a photonic network,
which is associated with a phase delay whose value depends on the
operating frequency.  The direction of propagation in each ring acts
as a two-fold degenerate degree of freedom, which can be thought of as
an analog of the electron spin in a quantum spin Hall
insulator\cite{RMPHasan}.  The primary ring in each unit cell is
coupled to its neighbors via waveguide loops \cite{hafezi}, shown in
Fig.~\ref{fig:couplings}(a) as a set of smaller rings.  If the
couplings have negligible internal backscattering, the inter-ring
coupling is ``spin'' conserving.  The clockwise and counter-clockwise
modes then form separate directed networks; the
network for clockwise modes is shown in Fig.~\ref{fig:couplings}(b).
The inter-link couplings, corresponding to the nodes of the network,
are described by unitary scattering matrices.

Propagation in such a network can be described by an evolution
operator \cite{klessemet1,HoChalker}.  Consider a unit cell of a
periodic network, such as the one shown in
Fig.~\ref{fig:couplings}(b).  For each cell, at lattice index $n$, we
can define a surface which is penetrated by $q$ input amplitudes
$\ket{a_n} \equiv [a_{1n}, \cdots, a_{qn}]$, and the same number of
output amplitudes $\ket{b_n} \equiv [b_{1n}, \cdots, b_{qn}]$.  The
input and output amplitudes are related by $S_{\textrm{int}} \ket{a_n}
= \ket{b_n}$, where $S_{\textrm{int}}$ is a unitary matrix describing
scattering from the interior of the designated surface.  As the
network is periodic, $S_{\textrm{int}}$ is independent of $n$.  We
will focus on the special case where the interior consists of
equal-length delay lines with phase delay $\phi$, as shown in
Fig.~\ref{fig:couplings}(b).  Then, with appropriate definitions of
$\ket{a}$ and $\ket{b}$,
\begin{equation}
  \ket{a_n} = e^{-i\phi} \, \ket{b_n}.
  \label{internal scattering}
\end{equation}
Furthermore, due to the connections between neighboring unit cells,
the amplitudes $\ket{b_n}$ leaving the surface of cell $n$ scatters
with those from other cells.  For Bloch modes, $\ket{a_{n}} =
\ket{a_k} e^{ik\cdot r_n}$ and $\ket{b_{n}} = \ket{b_k} e^{ik\cdot
  r_n}$, the inter-cell scattering can be described by
\begin{equation}
  S(k) \, \ket{b_k} = \ket{a_k},
  \label{external scattering}
\end{equation}
where $S(k)$ is unitary and is periodic in $k$ with the periodicity of
the Brillouin zone.  
The combination of Eqs.~(\ref{internal scattering})-(\ref{external
  scattering}) gives
\begin{equation}
  S(k) \, \ket{b_k} = e^{-i\phi} \,\ket{b_k}.
  \label{Sext equation}
\end{equation}
The eigenvectors of $S(k)$ are Bloch wave amplitudes, and the
arguments of the eigenvalues form a bandstructure $\phi(k)$.  The phase
delay $\phi$ is analogous to the band energy of a Bloch electron, or
the band frequency in a photonic crystal, apart from the fact that it
is an angle variable ($\phi \equiv \phi + 2\pi$).  Hereafter, we will
refer to $\phi$ as the ``quasi-energy''.

From the above description of a periodic network, we can see that the
modes of such a network are equivalent to the Floquet modes of a
periodically-driven lattice.  Suppose we have a lattice (having the same spatial dimension
as our network) whose Hamiltonian is periodic in time, with period
$T$.  Then Eq.~(\ref{Sext equation}) is the equation for a Floquet
state with state vector $\ket{b_k}$ and quasi-energy $\phi(k)/T$,
provided $S(k)$ is the time evolution operator over one period.
Explicitly,
\begin{equation}
  S(k) = \mathcal{T} \exp\left[-i\int_0^T dt \; H_k(t)\right],
\end{equation}
where $H_k(t)$ is some time-periodic reduced Hamiltonian and
$\mathcal{T}$ is the time-ordering operator.  (Except in special
cases, an explicit expression for $S(k)$ cannot be obtained from $H_k(t)$
or vice versa, but it can be computed numerically.)  The link between
network models and Floquet lattices has previously been pointed out
\cite{klessemet2,janssen,janssenbook}, but to our knowledge the consequences on
the band topology of network models has not been systematically
explored.

\section{Floquet band topology of network models}

Let us consider how the topology of a periodic network's bandstructure
might be characterized.  Following the usual topological
classification of band insulators \cite{Schnyder08,Schnyder09,Kitaev}, one
might take the matrix logarithm of Eq.~(\ref{Sext equation}) to obtain
an effective time-independent Hamiltonian, then look for topologically
non-trivial bands by computing topological band invariants (e.g.~the
Chern number for a 2D lattice without time-reversal
symmetry\cite{TKNN}).  However, doing so for the square lattice
network in the large-$\theta$ phase reveals that the
Chern number is zero despite the presence of topologically protected
``one-way'' edge states.  As discussed in Ref.~\onlinecite{Levin},
such ``anomalous Floquet insulator'' (AFI) behavior can arise in
Floquet bandstructures because the quasi-energy $\phi$ is an angle
variable.  At the topological transition, each band has simultaneous
Dirac band-crossing points with the band ``above'' and the band
``below'', modulo $2\pi$; these band-crossing points are respectively
associated with +1 and -1 Berry flux, so that the band has zero Chern
number on both sides of the transition.  In a static gapped
Hamiltonian system, the number of chiral edge states in a bulk gap can
be related to the sum of Chern numbers for all bands below the gap,
but this does not apply to Floquet systems since the quasi-energy
$\phi$ of a Floquet evolution operator is periodic and not bounded
below.

\begin{figure}
  \centering\includegraphics[width=0.45\textwidth]{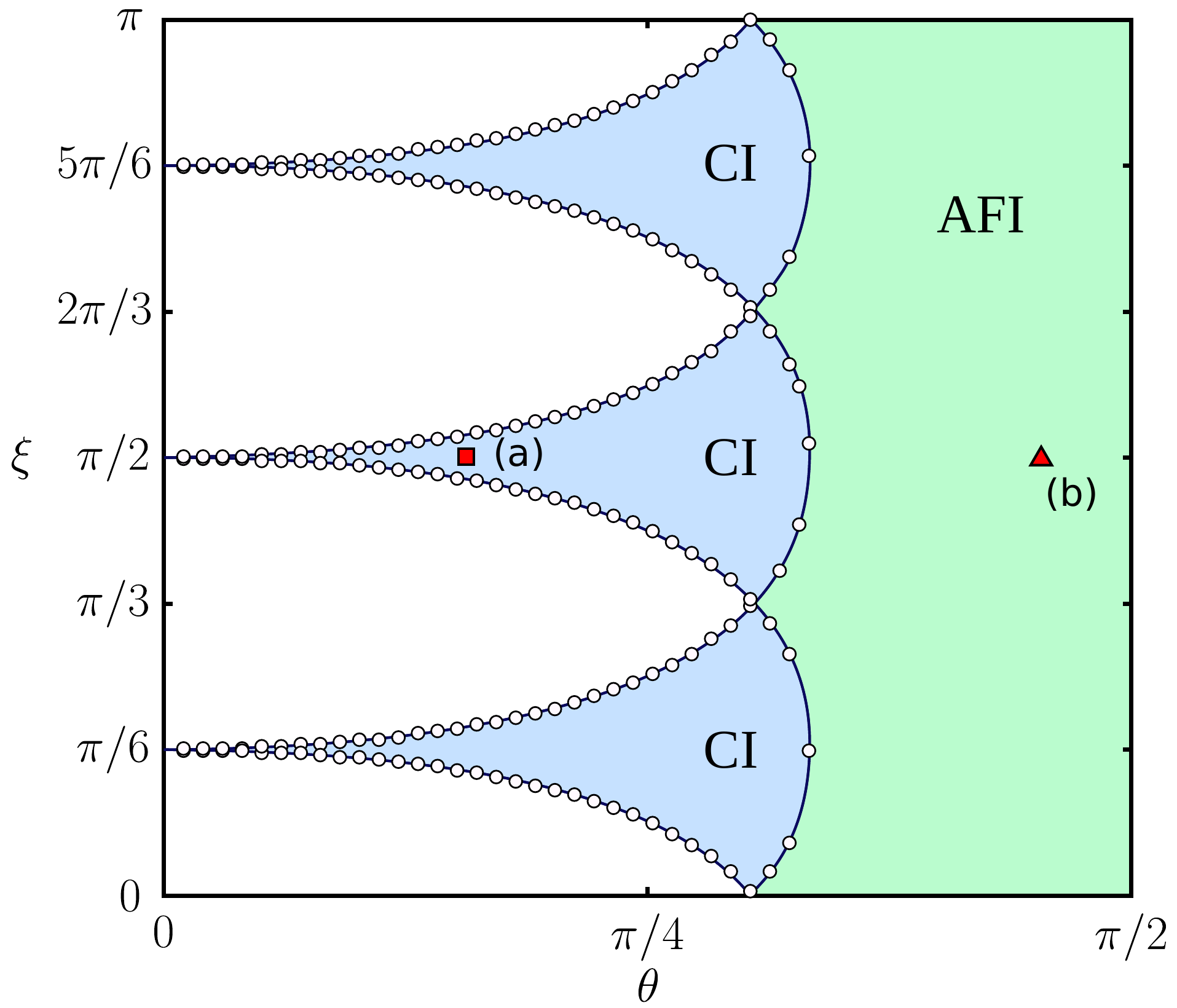}
   \caption{(color online) Phase diagram of a honeycomb network.
     The network is described in Appendix
     \ref{honeycomb appendix}; here we take coupling matrix parameters $\varphi=\chi=0$.  The phase boundaries are found by
     searching numerically for band crossings (dots).  The ``Chern
     insulator'' (CI) and ``anomalous Floquet insulator'' (AFI) phases
     are topologically non-trivial phases where the bands have
     non-zero and zero Chern number, respectively.  The unlabeled
     phases are conventional insulators.  The points labeled (a) and
     (b) indicate the parameters used for the projected band diagrams
     in Fig.~\ref{fig:hexbands}(a) and (b), respectively.}
   \label{fig:honphasediag}
\end{figure}

\begin{figure}
  \centering\includegraphics[width=0.35\textwidth]{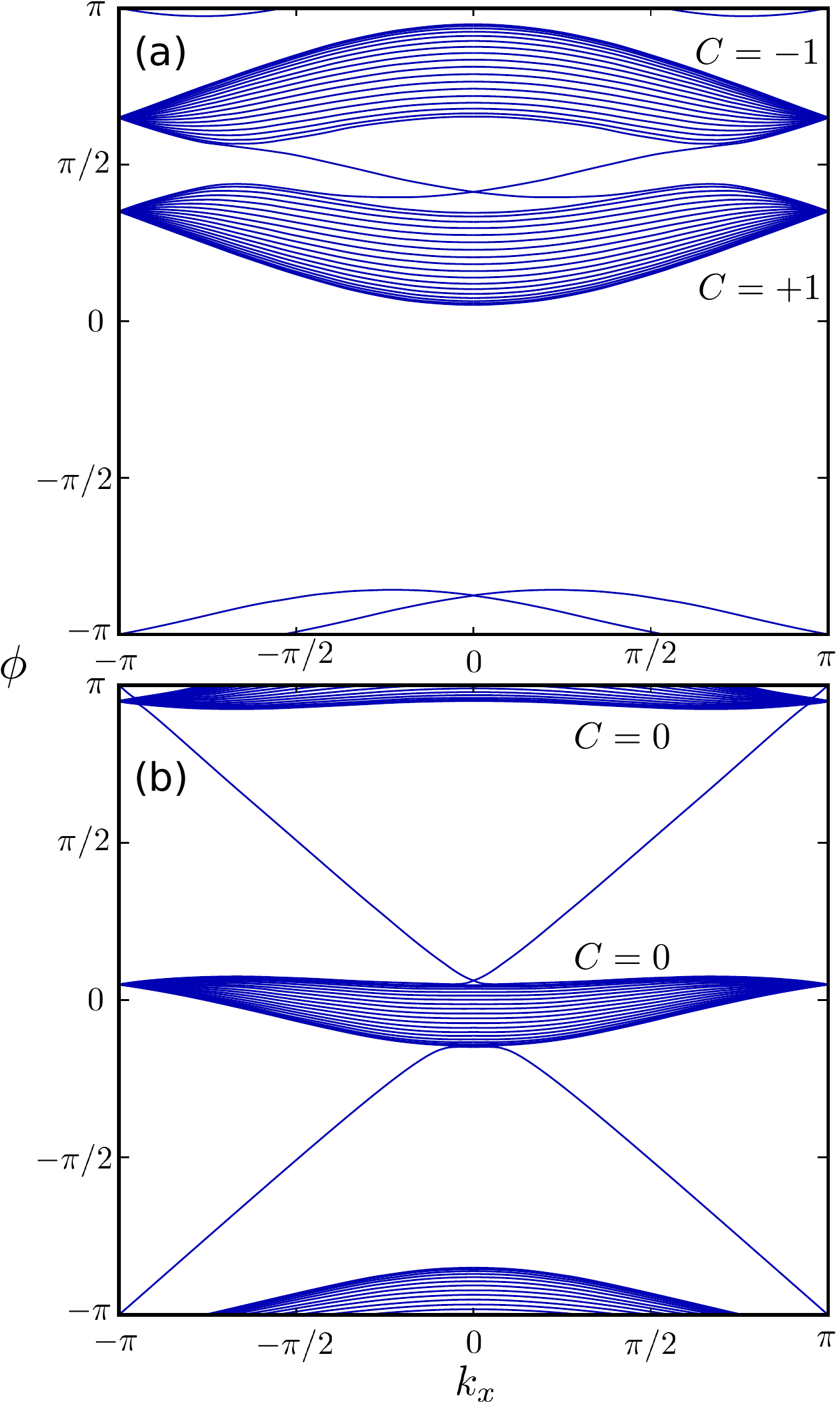}
  \caption{(color online) Projected quasi-energy bandstructures for
    the honeycomb network, in a strip geometry with width $N=20$ unit
    cells and zigzag edges.  The bands are computed from
    Eq.~(\ref{hexbands}); see Fig.~\ref{fig:hexnet} for a schematic of
    the network.  The coupling parameters are $\xi=\pi/2$,
    $\varphi=\chi=0$, and (a) $\theta=0.15 \pi$ (CI phase; upper figure),
    (b) $\theta=0.45 \pi$ (AFI phase; lower figure). The Chern number
    $C$ for each band is indicated.  These Chern numbers were computed
    from the momentum-space line integral of the Berry connection
    $\vect{\mathcal{A}}^{nn} (k) = -i\bra{nk}\vect{\nabla}_k\ket{nk}$,
    where $\ket{nk}$ is the $n$th Bloch eigenstate \cite{TKNN}.}
  \label{fig:hexbands}
\end{figure}

The square-lattice network has a rather simple phase diagram: it is an
AFI for values of the inter-ring coupling strength $\theta > \pi/4$,
and a conventional insulator otherwise, regardless of all other model
parameters.  However, more complicated behaviors can be observed in
other network models, such as networks based on a honeycomb lattice.
To our knowledge, such networks have not been studied previously,
partly because the network model literature was focused on the
Anderson transition, and the lattice geometry was not thought
to have a significant influence on properties such as the critical
exponent of the localization length\cite{ChalkerCo}.  The honeycomb
network, which is described in Appendix \ref{honeycomb appendix}, has
phases that depend on the inter-ring coupling $\theta$ as well as on
the parameters $\xi$ and $\varphi$, which describe the phase shifts
induced at the nodes [cf.~Eq.~(\ref{eq:Scat})-(\ref{eq:bandstruct})].  The phase
diagram for $\varphi = 0$ is shown in Fig.~\ref{fig:honphasediag}.
Unlike in the square lattice, topologically non-trivial phases exist
even for low values of $\theta$.  In these low-$\theta$ ``Chern
Insulator'' (CI) phases, the bands have non-zero Chern number, similar
to 2D systems with broken time-reversal symmetry \cite{haldane88}, and
the projected bandstructure exhibits topological edge states as shown
in Fig.~\ref{fig:hexbands}(a).  At larger values of $\theta$, the
system undergoes a transition from a CI phase to an AFI phase, where
all bands have zero Chern number and \emph{all} bandgaps are traversed
by topologically protected edge states\cite{Demler,Levin}, as shown in
Fig~\ref{fig:hexbands}(b).

As pointed out by Kitagawa \textit{et al.}, Floquet bandstructures can
be characterized by homotopy class-based topological invariants
\cite{Demler}, such as the ``$\nu_1$ invariants''
\begin{equation*}
  \frac{1}{2\pi}\int_{-\pi}^{\pi} dk_\mu
  \textrm{Tr}\left[S(k)^{-1} \; i\partial_{k_\mu} S(k) \right]
  \label{Demler invariant}
\end{equation*}
for $\mu = x,y$ in 2D.
In simple terms, these are the winding numbers for the quasi-energy
bands over their $[0,2\pi]$ domain, as $k_\mu$ is advanced through
$[0,2\pi]$.  They are non-zero in the AFI phase, where every bandgap
is topologically non-trivial and occupied by edge states; however, the
winding numbers are zero in CI phases where at least one of the
bandgaps is topologically trivial \cite{Demler}.  Subsequently, Rudner
\textit{et al.} have shown that the nontrivial topology of both the
AFI and CI phases can be
characterized by a bulk $\nu_3$ invariant
\cite{Levin}.  This invariant involves integrals over $k_x$ and $k_y$,
and over the time variable $t$.  In the context of network models,
there is no meaningful definition of the ``evolution operator'' for
intermediate $t$.  In practice, one can define any $S(k,t)$, such that
$S(k,T)$ is the evolution operator for the network; the choice is
non-unique but will not affect the value of $\nu_3$ thus obtained.

In the following section, we will investigate an alternative
topological characterization based on adiabatic pumping.  As we shall
see, the adiabatic pumping procedure is also capable of distinguishing
the AFI and CI phases, and it has the additional advantage of having a
natural physical interpretation for network models, which could be
useful for understanding the general class of Floquet bandstructures.

\section{Adiabatic pumping method and edge state invariants}

\begin{figure*}
  \centering\includegraphics[width=0.96\textwidth]{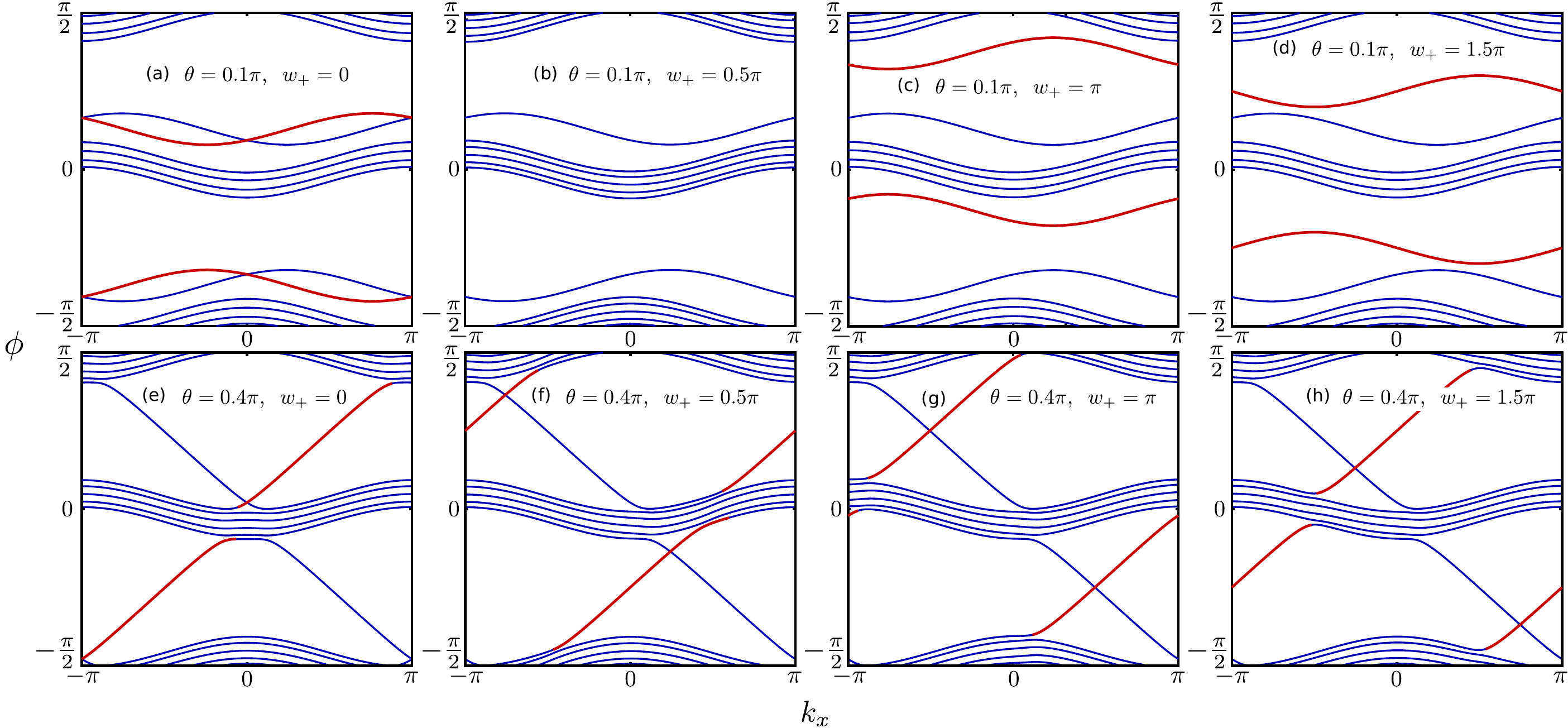}
  \caption{(color online) Projected bandstructures for the periodic
    square-lattice network of Fig.~\ref{fig:couplings}(c), with $N_y =
    6$ periods in the $y$ direction.  (a)-(d) show topologically
    trivial bandstructures ($\theta = 0.1\pi$, where $\theta$ is the
    inter-ring coupling strength \cite{Liang}), and
    (e)-(h) show topologically non-trivial bandstructures ($\theta =
    0.4\pi$).  Varying $w_+$, the angle variable controlling the upper
    edge, affects the edge states on the upper edge (highlighted in
    red).  The lower edge angle is fixed at $w_- = 0$, and the other
    coupling matrix parameters \cite{Liang} are $\varphi = \chi = 0$,
    $\xi = \pi/2$. }
  \label{fig:projected}
\end{figure*}

The adiabatic pumping method of characterizing topological systems was
originally introduced by Laughlin \cite{Laughlin}, and we will adapt
an elegant re-formulation of the Laughlin argument which was recently
given by Meidan \textit{et al.}~\cite{Brouwer}.  Working in the
context of static Hamiltonian systems, these authors imagined rolling
a 2D lattice into a cylinder by applying twisted boundary conditions
along one direction, attaching scattering leads to one cylinder edge,
and then calculating the eigenvalues of the scattering (reflection)
matrix.  As the twist angle is swept through $[0,2\pi]$, phase shifts
in the scattering eigenvalues can be related, via standard scattering
theory, to the number of resonances crossing the specified energy.
For mid-gap energies, scattering resonances correspond to edge states
of the isolated cylinder, which can be thus counted by the winding
numbers of the scattering matrix's eigenvalue spectrum \cite{Brouwer}.

A similar procedure can be carried out in a network model.  Let us
consider a two dimensional network, which is infinite in (say) the $x$
direction, and finite in the $y$ direction with $N_y$ periods.  For
convenience, we normalize the lattice spacings so the quasimomentum
$k_x$ becomes an angle variable.  The system can be regarded as a
supercell of $N_y$ unit cells, featuring twisted boundary conditions
along the $x$ boundaries with twist angle $k_x$.  Following the
discussion in Section \ref{sect:Floquet maps}, we can designate a
scattering surface for this supercell, consisting of the union of the
scattering surfaces for the individual unit cells.  This is shown in
Fig.~\ref{fig:couplings}(c) for the simple square-lattice network.
The inputs entering this supercell surface are $\ket{a} = [\ket{a_1},
  \cdots \ket{a}_{N_y}]$, and the output amplitudes are $\ket{b} =
[\ket{b_1}, \cdots \ket{b_{N_y}}]$.  The scattering from the interior
of the surface gives $\ket{a} = e^{-i\phi}\, \ket{b}$.  As for the
scattering from the exterior of the surface back into the interior,
that depends on the inter-cell connections (which are assumed
constant), and on $k_x$ (due to scattering across the $x$ boundaries).
There is one more set of constraints which must also be specified: the
relations between the input and output amplitudes penetrating the
scattering surface along the $y$ boundaries of the supercell.  As
depicted in Fig.~\ref{fig:couplings}(c), we denote these ``edge
amplitudes'' by $\ket{a_\pm}$ and $\ket{b_\pm}$, with the $\pm$
subscripts indicating the upper and lower edges.  Let the number of
edge amplitudes on each edge be $n_{\perp}$.  In general, we have
\begin{equation}
  S_\perp \begin{bmatrix}\,\ket{b_+}\, \\ \,\ket{b_-}\,\end{bmatrix}
    = \begin{bmatrix}\,\ket{a_+}\, \\ \,\ket{a_-}\,\end{bmatrix}
\end{equation}
for some some $2n_\perp\times2n_\perp$ unitary matrix $S_\perp$.  From
this, we can construct an exterior scattering matrix for the
super-cell, $S_\textrm{sc}$, such that
\begin{equation}
  S_\textrm{sc}(k_x,S_\perp) \, \ket{b}= e^{-i\phi} \,\ket{b}.
  \label{supercell}
\end{equation}
We are free to specify $S_\perp$, and it is useful to consider a case
where the upper and lower boundaries are ``disconnected''.
Specifically,
\begin{equation}
  S_\perp(w_+,w_-) = \begin{bmatrix}e^{iw_+} I & 0 \\ 0 & e^{iw_-} I
  \end{bmatrix}.
  \label{sperp}
\end{equation}
The values of $\phi(k_x)$ obtained from
Eqs.~(\ref{supercell})-(\ref{sperp}) form a projected quasi-energy
bandstructure for the semi-infinite lattice of width $N_y$, with the
set of $2n_\perp$ edge angles, $\{w_\pm\}$, acting as tunable edge
conditions.

The edge angles $w_\pm$ can be used to define topological invariants.
Suppose we keep $w_-$ fixed and consider only variations in $w_+$.
For any $\phi, k_x \in [0,2\pi]$, there must be exactly $n_\perp$
values of $w_+ \in [0,2\pi]$ consistent with
Eqs.~(\ref{supercell})-(\ref{sperp}); in physical terms, by specifying
$\phi$ and $k_x$ (as well as fixing $w_-$ and other network parameters
entering into $S_\textrm{sc}$), we have defined an $n_\perp$-channel
scattering problem, and the input amplitudes $\ket{a_+}$ and output
amplitudes $\ket{b_+}$ for the scatterer must be related by some
unitary reflection matrix whose eigenvalues are $e^{iw_+}$.  Let us
fix a value for the quasi-energy $\phi$ which lies in a bulk bandgap,
and consider the $n_\perp$-valued function $w_+(k_x)$, which must come
back to itself (modulo $2\pi$) as $k_x$ is advanced over $[0,2\pi]$.
Each value of $w_+$ corresponds to a separate projected bandstructure,
but within each gap only the dispersion curves for edge states
localized to the upper edge can vary, since $w_+$ cannot affect the
lower edge.  As a result, the winding number of $w_+(k_x)$ counts the
net (forward minus backward) number of upper edge states in the
specified bandgap.

\begin{figure}
  \centering\includegraphics[width=0.44\textwidth]{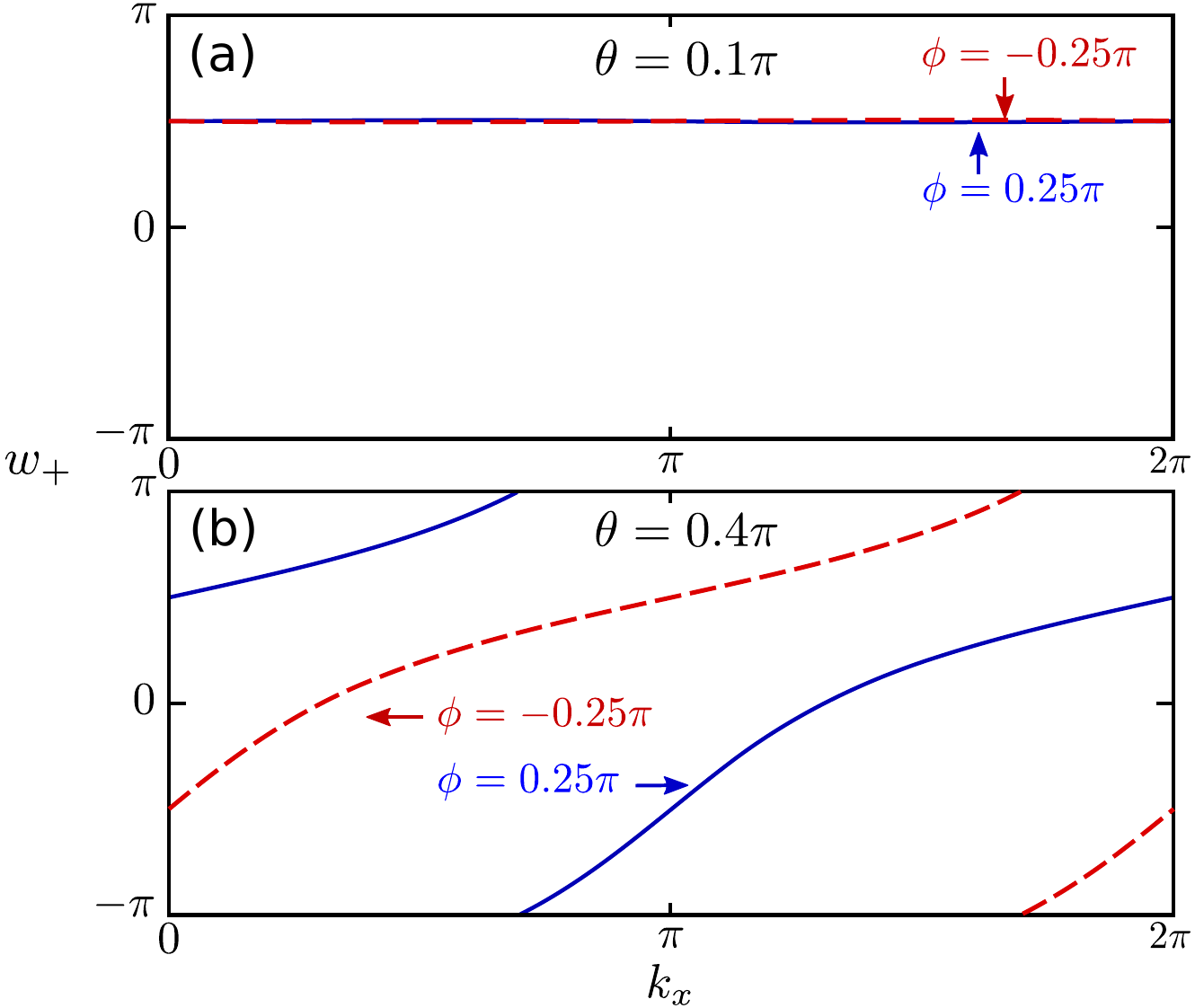}
  \caption{(color online) Plots of the edge angle $w_+$
    versus $k_x$, in the square-lattice network with width $N_y = 6$.
    (a) In the conventional insulator phase ($\theta = 0.1\pi$), the
    winding numbers are zero; (b) in the AFI phase ($\theta =
    0.4\pi$), the winding numbers are +1.  In both cases, plots are
    given for $\phi = \pi/4$ and $\phi = -\pi/4$, which lie in two
    different band gaps (see Fig.~\ref{fig:projected}).  In all cases,
    $w_- = 0$ and all the other parameters are the same as in
    Fig.~\ref{fig:couplings}.  }
  \label{fig:winding}
\end{figure}

To illustrate the above discussion, consider the previously-discussed
square-lattice network, for which $n_\perp = 1$ (i.e., $w_+(k_x)$ is
single-valued).  Projected bandstructures for this network are shown
in Fig.~\ref{fig:projected}; for details of the calculation, see
Appendix \ref{square appendix}.  In the conventional insulator phase,
corresponding to Figs.~\ref{fig:projected}(a)-(d), $w_+(k_x)$ has zero
winding number in each gap, as shown in Fig.~\ref{fig:winding}(a).
Note, however, that Fig.~\ref{fig:winding}(a) also shows that there
are certain values of $w_+$ for which upper edge states \textit{do}
exist.  In the projected bandstructure, these take the form of
isolated bands of \textit{two-way} edge states which are ``pumped''
downwards across each gap during each cycle of $w_+$.  

In the AFI phase, $w_+(k_x)$ has winding number +1 in each gap, as
shown in Fig.~\ref{fig:winding}(b).  The projected bandstructures,
shown in Figs.~\ref{fig:projected}(e)-(h), exhibit one-way edge states
spanning each gap.  Each band of edge states ``winds'' across the
Brillouin zone during one cycle of $w_+$, with the overall effect of
pumping one band down across each gap during one cycle of $w_+$, like
in the conventional insulator phase.  Each gap also has a band of edge
states that is invariant in $w_+$, corresponding to states localized
on the lower edge.

We expect this to be the generic effect of adiabatic pumping on
quasi-energy bandstructures. Because $w_+$ is a well-defined function of $k_x$,
winding $w_+$ by $2\pi$ has the effect of transporting a band of
edge states across each gap.  This transport occurs even for conventional
(topologically trivial) bandgaps, in the form of a band of two-way edge states.
The bandstructure as a whole returns to itself over one such cycle, which
is possible since the quasi-energy is an angle variable.

In the honeycomb network, the conventional insulator and AFI phases
behave in the same way as for the square-lattice network.  In the CI phase, each
cycle of $k_x$ transports a band of two-way edge states down across
the topologically trivial gap (where $w_+(k_x)$ has zero winding
number), while simultaneously winding the one-way edge states in the
topological gap (where $w_+(k_x)$ has winding number +1).

\begin{figure}
  \centering\includegraphics[width=0.41\textwidth]{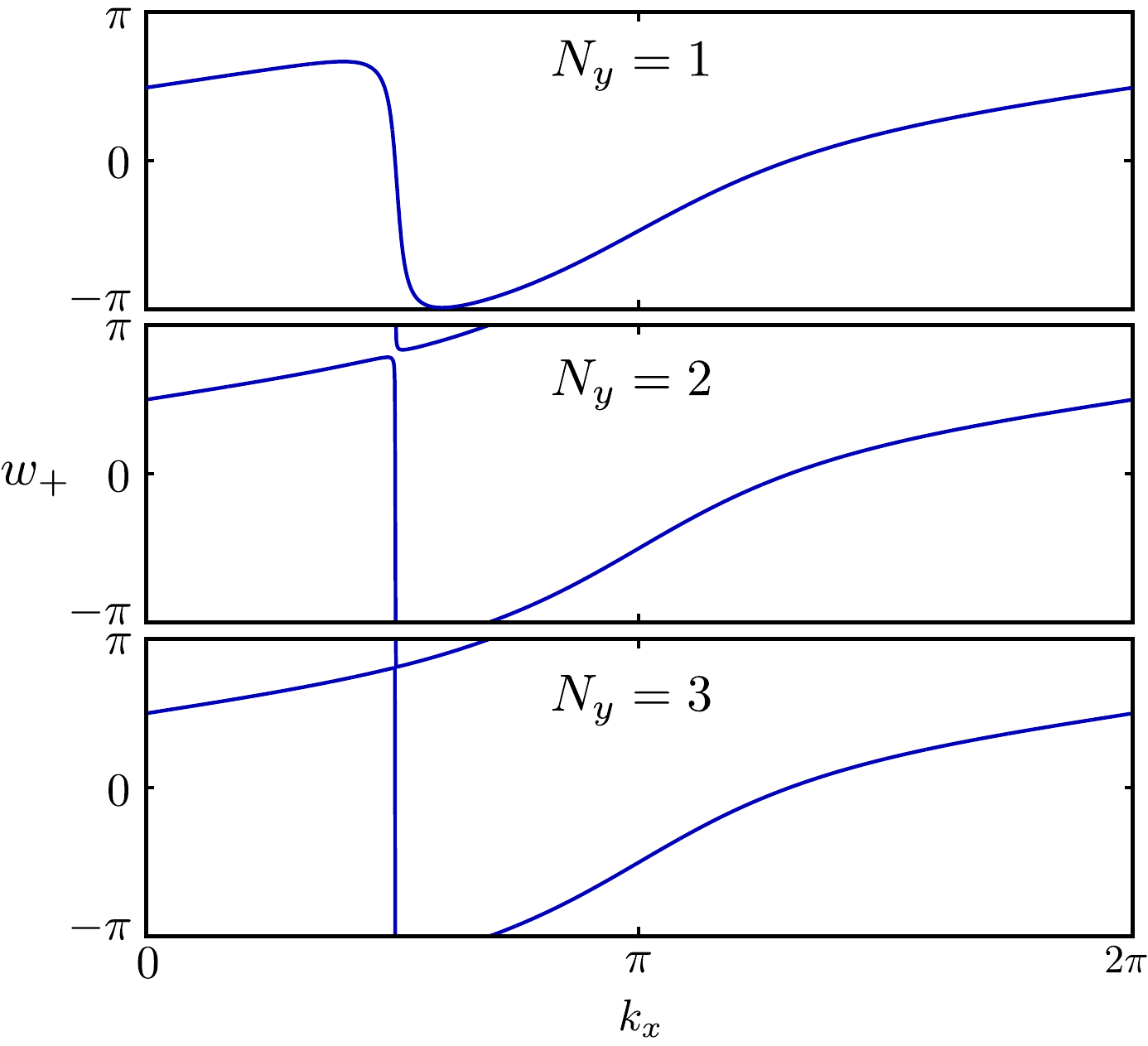}
  \caption{(color online) Plots of $w_+$ versus $k_x$ for small values
    of $N_y$, showing the emergence of a non-zero winding number.  For
    all three plots, we use $\phi = 0.25\pi$ and $\theta = 0.4\pi$,
    corresponding to a mid-gap quasi-energy in the AFI phase.  All other parameters are as in
    Fig.~\ref{fig:projected}.  For $N_y > 1$, an anti-crossing
    develops near $k_x \sim \pi/2$, coinciding with the dispersion
    curve for the lower edge states in the projected band diagram.
    The width of this anti-crossing goes rapidly to zero with $N_y$,
    and the rest of the curve acquires a non-zero winding
    number. }
  \label{fig:n}
\end{figure}

The relation of the winding number of $w_+(k_x)$ to the edge states
relies on the assumption that the upper edge angles have no effect on
the lower edge states.  Hence, $\phi$ must to be chosen within a
bandgap, and the width $N_y$ must be sufficiently large (compared to
the edge state penetration depth).  This is demonstrated in
Fig.~\ref{fig:n}, where we plot $w_+(k_x)$ using $N_y = 1, 2, 3$, for
the square-lattice network in the AFI phase.  For $N_y = 1$, we
observe that $w_+(k_x)$ has zero winding number.  As $N_y$ is
increased, the curve develops an anti-crossing, occurring at a value
of $k_x$ coinciding with the quasimomentum of an edge state localized
to the lower edge (for the specified value of $\phi$).  For
sufficiently large $N_y$, the lower edge state is independent of
$w_+$, so the anti-crossing narrows into a numerically-undetectable
vertical line.  Because the anti-crossing is associated with a $-1$
winding number, the remainder of the $w_+(k_x)$ curve acquires $+1$
winding.

\section{Discussion}

In this paper, we have discussed the relationships between photonic
resonator lattices, Chalker-Coddington network models, and Floquet
topological insulators.  Within the emerging field of topological
photonics, these analogies may provide insights for realizing new
topological phases.  For example, some years ago Chalker and Dohmen
\cite{ChalkerDoh} studied a hypothetical three-dimensional network
consisting of weakly-coupled 2D stacked layers of CC networks (a
configuration reminiscent of a 3D weak topological
insulator\cite{RMPHasan}).  Photonic lattice analogs of such 3D networks may
be realizable, possibly at microwave frequencies
for ease of fabrication.  Furthermore, as discussed in the
introduction, a photonic Floquet topological insulator has recently
been realized \cite{Szameit}, in which the 2D bands were shown to possess non-zero
Chern numbers.  It would be interesting to analyze this or a similar
system using the scattering formalism of a network model, with the aim
of realizing an AFI phase where topologically-protected edge states
are present despite all bands having zero Chern number.  (A photonic
AFI-like phase has previously been realized in 1D
\cite{KitagawaNatCom}.)

We have restricted our attentions to \emph{directed} network models.
In the photonic context, this means considering the flow of light in a
single direction within the waveguides, and assuming no backscattering
into time-reversed modes.  Apart from this restriction, there are no
further symmetry requirements on the coupling matrices.  The two
possible directions of propagation through the network are analogous
to two decoupled spin sectors in a 2D quantum spin Hall insulator.
However, in the electronic case a topological phase can exist even in
the presence of spin-mixing: the ${\mathbb Z} _2$ topological
insulator.  This relies on the fact that edge states cannot be
backscattered by time-reversal symmetric perturbations due to the
particular nature of \textit{fermionic} time-reversal symmetric $S$
matrices \cite{RMPHasan}.  Indeed, the CC network model concept has
been generalized to study quantum spin Hall insulators by imposing
fermionic time-reversal symmetries on the links and nodes
\cite{obuse1,obuse2,ryu}.  However, \textit{bosonic} edge states are not
protected from backscattering by time-reversal symmetric
perturbations, so topologically non-trivial behavior can only occur if
mixing into time-reversed modes is negligible.  This is an important limitation of
photonic topological insulators, but not necessarily a fatal one,
since such mixing processes can often be engineered away.

We have also, in this paper, considered translationally periodic
systems.  It would be interesting to return to the original motivation
for introducing network models, which was to study disorder-induced
Anderson transitions in a
2D electron gas \cite{ChalkerCo}.  In the photonic context, Anderson
\textit{localization} of light has been observed in 1D and 2D
\cite{Segev,Segevreview}.  However, there is no Anderson
\textit{transition} in such systems, since they map onto time-reversal
symmetric electron gases for which localization is marginal in 2D \cite{localizationRMP}.  By
contrast, an Anderson transition does exist in 2D disordered quantum
Hall systems, tied to the phenomenon of classical percolation
\cite{ChalkerCo}.  Random photonic networks might thus manifest a
photonic localization-delocalization transition, which has not yet
been observed.

\begin{acknowledgments}
We thank M.~Rechtsman, A.~Szameit, M.~Hafezi, and G.~Q.~Liang for
helpful comments.  This research was supported by the Singapore
National Research Foundation under grant No.~NRFF2012-02.
\end{acknowledgments}

\appendix

\setcounter{MaxMatrixCols}{30} 

\section{Bandstructure of a square lattice network}
\label{square appendix}


Fig.~\ref{fig:couplings}(b) shows a unit cell of the square lattice
network, which consists of two nodes with coupling relations
 \begin{align}
   S_x \begin{bmatrix}b_{1,n} \\ b_{3,n+x} \end{bmatrix}
   &= \begin{bmatrix} a_{4,n+x} \\ a_{2,n}  \end{bmatrix}, \label{Sx} \\
   S_y \begin{bmatrix} b_{4,n} \\ b_{2,n+y} \end{bmatrix}
   &= \begin{bmatrix} a_{3,n+y} \\ a_{1,n}  \end{bmatrix}, \label{Sy}
 \end{align}
 where
 \begin{equation}
	 S_\mu = \begin{bmatrix} r_\mu & t'_\mu \\ t_\mu & r'_\mu \end{bmatrix}.
\end{equation}
Using the relations between wave amplitudes $a_{i,n}$ and $b_{i,n}$
coming from link phases, we can write Eq.~(\ref{Sy}) as
\begin{equation}
 S_y \begin{bmatrix} a_{4,n} \\ a_{2,n+y} \end{bmatrix} = \begin{bmatrix} b_{3,n+y} \\ b_{1,n}  \end{bmatrix} e^{-2 i \phi}.
 \label{Sy1}
\end{equation}
From the translational invariance of the network strip in the $x$
direction, wave amplitudes in Eq.~(\ref{Sx}) can be written in the
Bloch form to obtain:
\begin{equation}
    S_x \begin{bmatrix}b_{1,n} \\ b_{3,n} e^{i k_x} \end{bmatrix} = \begin{bmatrix} a_{4,n} e^{i k_x} \\ a_{2,n}  \end{bmatrix}.
    \label{Sx1}
\end{equation}
By reordering the terms in (\ref{Sy1})-(\ref{Sx1}), one obtains
\begin{align}
  S'_x (k_x)  \begin{bmatrix} b_{3,n} \\ b_{1,n} \end{bmatrix}
  &= \begin{bmatrix}a_{2,n} \\ a_{4,n}  \end{bmatrix},\label{Sxy1}\\
  S'_y \begin{bmatrix}a_{4,n} \\ a_{2,n+y} \end{bmatrix}
  &= \begin{bmatrix}b_{1,n} \\ b_{3,n+y}  \end{bmatrix} e^{-2 i \phi}.
  \label{Sxy2}
\end{align}

In order to obtain the bandstructure of the square lattice network in
the strip geometry, we need to construct a scattering matrix for the
super-cell, $S_\textrm{sc}$, defined in Fig.~\ref{fig:couplings}(c).
This obeys
\begin{equation}
 S_\textrm{sc} (k_x,w_+,w_-) \ket{b} = e^{- i \phi T} \ket{b},
 \label{Ssc}
\end{equation}
where $\ket{b}$ is a wave amplitude vector, and the angles $w_+$ and
$w_-$ set the boundary conditions at the strip edges such that
(cf. Fig.~\ref{fig:couplings}):
\begin{align}
e^{i w_-} b_{2,1}  &= a_{3,1},\\
e^{i w_+}  b_{4,N_y}  &= a_{1,N_y},
  \end{align}
or, equivalently,
\begin{align}
e^{i w_-} a_{2,1}  &= b_{3,1} e^{-2 i \phi},\label{wangles1}\\
e^{i w_+}  a_{4,N_y}  &= b_{1,N_y} e^{-2 i \phi}.
 \label{wangles2}
  \end{align}
Finally, using Eqs.~(\ref{Sxy1}, \ref{Sxy2}, \ref{wangles1}, \ref{wangles2}) one can construct the $2 N_y \times 2 N_y$ matrices $M_A$ and $M_B$ such that
\begin{align}
 M_A (w_+, w_-) &= \begin{bmatrix}
        e^{i w_-} &      &       &      & \\
                  & S'_y &       &      & \\
                  &      & \ddots&      & \\
                  &      &       & S'_y & \\
                  &      &       &      & e^{i w_+}
       \end{bmatrix},\\
       M_B (k_x) &= \begin{bmatrix}
        S'_x &        &       \\
             & \ddots &       \\
             &        & S'_x   
       \end{bmatrix},
\end{align}
to obtain
\begin{equation}
M_A (w_+,w_-) M_B(k_x) \ket{b'} = e^{-2 i \phi} \ket{b'},
\end{equation}
where $\ket{b'} = [ b_{3,1}, b_{1,1}, \cdots, b_{3, N_y}, b_{1, N_y}]$, which is similar to Eq.~(\ref{Ssc}) with $T=2$.

\section{Bandstructure and phase diagram of honeycomb network}
\label{honeycomb appendix}

\begin{figure}
  \centering\includegraphics[width=3.7cm]{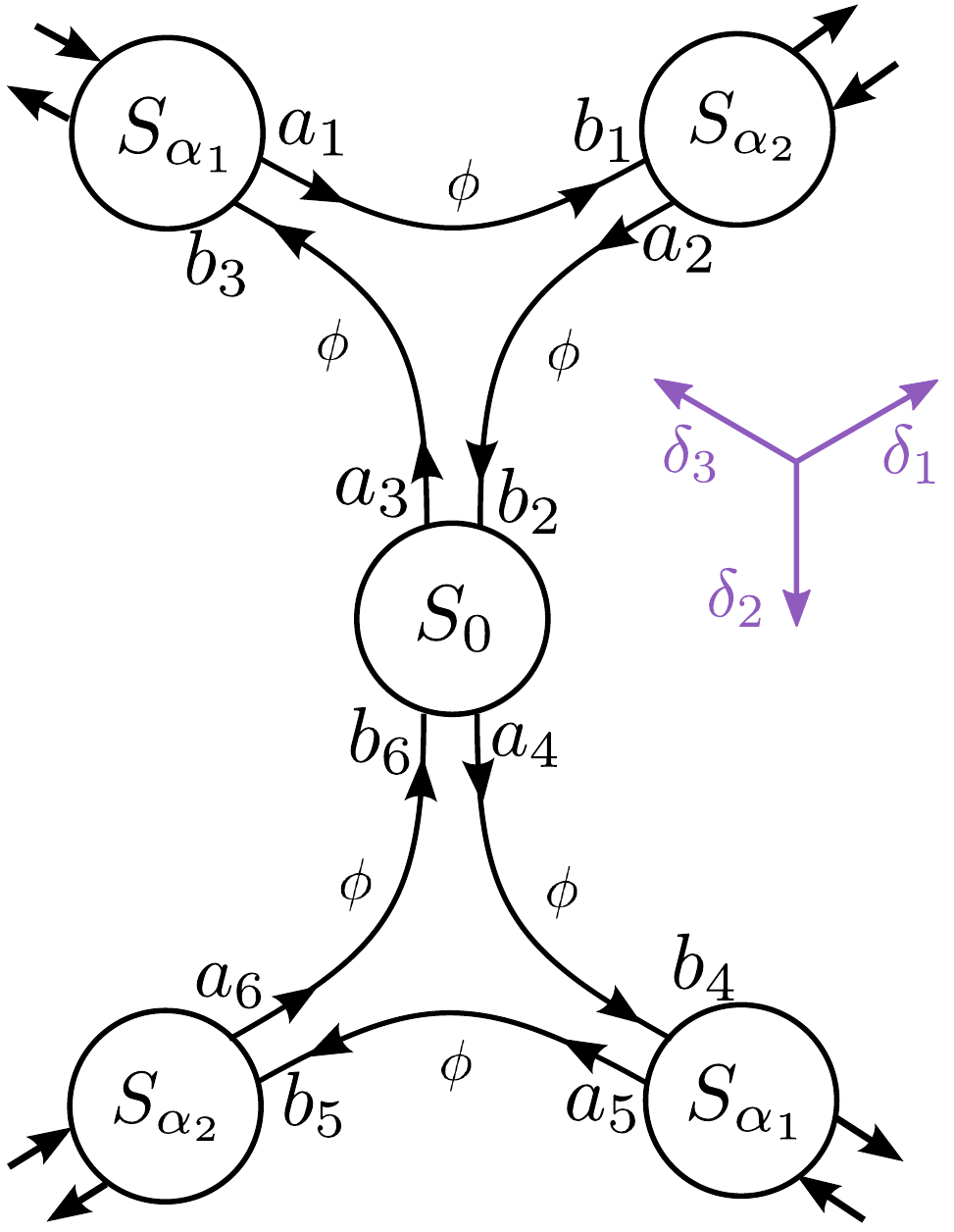}
  \caption{(color online) Schematic of a unit cell in a
    two-dimensional honeycomb periodic network.  
    }
  \label{fig:hexnet}
  \end{figure}

The honeycomb network unit cell is represented in
Fig.~\ref{fig:hexnet}, with the corresponding wave amplitudes.  We
define the scattering relations at the nodes of the network such that
the first (resp. second) reflection block of the S-matrix describes
the hopping in the $+\delta_i$ ($-\delta_i$) direction, where
$i=1,2,3$.
This gives
\begin{align}
   S_{\alpha_2} \begin{bmatrix}b_{1,n} \\ b_{5,n+\alpha_2} \end{bmatrix}
   &= \begin{bmatrix} a_{6,n+\alpha_2} \\ a_{2,n}  \end{bmatrix}, \label{Sa2} \\
   S_{\alpha_1} \begin{bmatrix} b_{3,n} \\ b_{4,n+\alpha_1} \end{bmatrix}
   &= \begin{bmatrix} a_{5,n+\alpha_1} \\ a_{1,n}  \end{bmatrix}, \label{Sa1} \\
   S_0 \begin{bmatrix} b_{2,n} \\ b_{6,n} \end{bmatrix}
   &= \begin{bmatrix} a_{4,n} \\ a_{3,n}  \end{bmatrix}. \label{S0}
 \end{align}
Using the phase relations on network links, we can rewrite
Eqs.~(\ref{Sa2})-(\ref{S0}) as
\begin{align}
   S_{\alpha_2} \begin{bmatrix}b_{1,n} \\ b_{5,n+\alpha_2} \end{bmatrix}
   &= \begin{bmatrix} a_{6,n+\alpha_2}\\ a_{2,n}  \end{bmatrix}, \label{Sa2b} \\
 S_{\alpha_1} \begin{bmatrix} a_{3,n} \\ a_{4,n+\alpha_1} \end{bmatrix}
   &= \begin{bmatrix} b_{5,n+\alpha_1} \\ b_{1,n}  \end{bmatrix} e^{-2 i \phi},\label{Sa1b}\\
   S_0 \begin{bmatrix} a_{2,n} \\ a_{6,n} \end{bmatrix}
   &= \begin{bmatrix} a_{4,n} \\ a_{3,n}  \end{bmatrix} e^{-i\phi}.\label{S0b}
\end{align}
We can now use Bloch's theorem, taking the honeycomb network to be
translationally invariant in the $\vect{\alpha_2}$ direction (which
yields zig-zag edges).  Eqs.~(\ref{Sa1b}) and (\ref{Sa2b}) become
\begin{align}
 S'_{\alpha_1} \begin{bmatrix} a_{3,n} \\ a_{4,n+\alpha_1} \end{bmatrix}
   &= \begin{bmatrix} b_{1,n} \\ b_{5,n+\alpha_1}  \end{bmatrix} e^{-2 i \phi}\\
  S'_{\alpha_2}(k) \begin{bmatrix}b_{5,n} \\ b_{1,n}  \end{bmatrix}
   &= \begin{bmatrix} a_{2,n} \\ a_{6,n}  \end{bmatrix}.
\end{align}
We have the edge angles relations as
\begin{align}
e^{i w_-} b_{4,1}  &= a_{5,1},\\
e^{i w_+}  b_{3,N_{\alpha_1}}  &= a_{1,N_{\alpha_1}},
  \end{align}
or, equivalently,
\begin{align}
e^{i w_-} a_{4,1}  &= b_{5,1} e^{-2 i \phi},\\
e^{i w_+}  a_{3,N_{\alpha_1}}  &= b_{1,N_{\alpha_1}} e^{-2 i \phi}.
  \end{align}
One can thus construct the following matrices:
  \begin{align}
 M_A (w_+, w_-) &=  \begin{bmatrix}
		    e^{i w_-} &      &       &      & \\
			      & S'_{\alpha_1} &       &      & \\
			      &      & \ddots&      & \\
			      &      &       & S'_{\alpha_1} & \\
			      &      &       &      & e^{i w_+}
		    \end{bmatrix},\\
      M_B &= 
       \begin{bmatrix}
        S_0 &        &       \\
             & \ddots &       \\
             &        & S_0   
       \end{bmatrix},\\
     M_C (k) &= \begin{bmatrix}
        S'_{\alpha_2} &        &       \\
		      & \ddots &       \\
		      &        & S'_{\alpha_2}   
       \end{bmatrix},
\end{align}
such that
\begin{equation}
M_A (w_+,w_-) M_B M_C(k) \ket{b'} = e^{-3 i \phi} \ket{b'},
\label{hexbands}
\end{equation}
where $\ket{b'} = [ b_{5,1}, b_{1,1}, \cdots, b_{5, N_y}, b_{1, N_y}]$.

Setting $S_{\alpha_1}=S_{\alpha_2}=S_{0}$ for simplicity, we can use
the phase delay relations between amplitudes on the honeycomb network
to eliminate $a_1, a_3, a_4, a_5$ and $b_2, b_6$.  Then
Eqs.~(\ref{Sa2})-(\ref{S0}) reduce to
\begin{subequations}
\label{eq:Smatgrp}
\begin{align}
r b_1 + t' b_5 e^{i \vect{k}\cdot \vect{\alpha_2}} &= a_6 e^{i \vect{k}\cdot \vect{\alpha_2}} \label{eq:Smatgrp1}\\
t b_1 + r' b_5 e^{i \vect{k}\cdot \vect{\alpha_2}} &= a_2\label{eq:Smatgrp2}\\
r b_3 + t' b_4 e^{i \vect{k}\cdot \vect{\alpha_1}} &= b_5  e^{-i \phi} e^{i \vect{k}\cdot \vect{\alpha_1}}\label{eq:Smatgrp3}\\
t b_3 + r' b_4 e^{i \vect{k}\cdot \vect{\alpha_1}} &= b_1 e^{-i \phi}\label{eq:Smatgrp4}\\
r a_2 e^{i \phi} + t' a_6 e^{i \phi} &= b_4 e^{- i \phi}\label{eq:Smatgrp5}\\
t a_2 e^{i \phi} + r' a_6 e^{i \phi}  &= b_3 e^{-i \phi}. \label{eq:Smatgrp6}
\end{align}
\end{subequations}
Here we have used Bloch's theorem, e.g.~$b_{5,n+\alpha_2} = b_{5} e^{i
  \vect{k}\cdot \vect{a_2}}$ (discarding the index $n$).  With
Eqs.~(\ref{eq:Smatgrp5}) and (\ref{eq:Smatgrp6}), we can eliminate
$b_3$ and $b_4$:
\begin{subequations}
\label{eq:Smat2grp}
\begin{align}
r b_1 + t' b_5 e^{i \vect{k}\cdot \vect{\alpha_2}} &= a_6 e^{i \vect{k}\cdot \vect{\alpha_2}} \label{eq:Smat2grp1}\\
t b_1 + r' b_5 e^{i \vect{k}\cdot \vect{\alpha_2}} &= a_2\label{eq:Smat2grp2}\\
\begin{split}
r [t a_2 e^{i 2\phi} + r' a_6 e^{i 2\phi}]\, +\, & t'e^{i \vect{k}\cdot \vect{\alpha_1}} [r a_2 e^{i 2\phi} + t' a_6 e^{i 2\phi}] \\
&= b_5  e^{-i \phi} e^{i \vect{k}\cdot \vect{\alpha_1}}
\end{split} \label{eq:Smat2grp3}\\
\begin{split}
t [t a_2 e^{i 2\phi} + r' a_6 e^{i 2\phi}]\, +\,& r'e^{i \vect{k}\cdot \vect{\alpha_1}} [r a_2 e^{i 2\phi} + t' a_6 e^{i 2\phi}] \\
&= b_1 e^{-i \phi}.
\end{split}  \label{eq:Smat2grp4}
\end{align}
\end{subequations}
Finally, using Eqs.~(\ref{eq:Smat2grp1}) and (\ref{eq:Smat2grp2}), we
eliminate $a_2$ and $a_6$ to obtain:
\begin{widetext}
\begin{multline}\label{eq:Smat3grp2}
 b_1 [r t^2 e^{i 2\phi} + r^2 r' e^{-i \vect{k}\cdot \vect{\alpha_2}} e^{i 2\phi} + r t t' e^{i \vect{k}\cdot \vect{\alpha_1}} e^{i 2\phi} 
 + r t'^2 e^{i \vect{k}\cdot (\vect{a_1}-\vect{\alpha_2})}e^{i 2\phi}] \\ 
= b_5 [e^{-i \phi} e^{i \vect{k}\cdot \vect{\alpha_1}} - r r' t e^{i \vect{k}\cdot \vect{\alpha_2}} e^{i 2\phi} - r r' t' e^{i 2\phi} - r r' t' e^{i \vect{k}\cdot (\vect{\alpha_1}+\vect{\alpha_2})} e^{i 2\phi} - t'^3 e^{i \vect{k}\cdot \vect{\alpha_1}} e^{i 2\phi}],
\end{multline}
and
\begin{multline}
b_5 [ r' t^2  e^{i \vect{k}\cdot \vect{\alpha_2}} e^{i 2\phi} + r' t t' e^{i 2\phi} + r r'^2 e^{i \vect{k}\cdot (\vect{\alpha_1}+\vect{\alpha_2})} e^{i 2\phi} + r' t'^2 e^{i \vect{k}\cdot \vect{\alpha_1}} e^{i 2\phi}]\\
= b_1 [e^{-i \phi} - t^3 e^{i 2\phi} - r r' t e^{-i \vect{k}\cdot \vect{\alpha_2}} e^{i 2\phi} - r r' t e^{i \vect{k}\cdot \vect{\alpha_1}} e^{i 2\phi} - r r' t' e^{i \vect{k}\cdot (\vect{\alpha_1}-\vect{\alpha_2})} e^{i 2\phi}].
\end{multline}
After simplification, this yields:
\begin{multline}
 e^{i 6\phi} (r r' - t t')^3 +
 e^{i 3\phi} \left\lbrace t^3 + t'^3 + r r' \left[ t e^{i \vect{k}\cdot \vect{\alpha_1}} +t' e^{-i \vect{k}\cdot \vect{\alpha_1}} + t' e^{i \vect{k}\cdot \vect{\alpha_2}} + t e^{-i \vect{k}\cdot \vect{\alpha_2}} \right. \right.\\
 \qquad \left. \left.{} + t' e^{i \vect{k}\cdot (\vect{\alpha_1}-\vect{\alpha_2})} + t e^{i \vect{k}\cdot (\vect{\alpha_2}-\vect{\alpha_1})} \right] \right\rbrace - 1 =0.
\end{multline}
Using the following parameterization for the $2 \times 2$ unitary S-matrix\cite{Liang}:
\begin{equation}
 S = \begin{bmatrix} r = \sin \theta e^{i \chi}  & ~ t'= - \cos \theta e^{i (\varphi - \xi)} \\ t= \cos \theta e^{i \xi} & r'= \sin \theta e^{i (\varphi - \chi )}\end{bmatrix},
 \label{eq:Scat}
\end{equation}
and the hexagonal lattice vectors $ \vect{\alpha_1} = \frac{3}{2}
\vect{x} + \frac{\sqrt{3}}{2} \vect{y}$ , $ \vect{\alpha_2}
=\frac{3}{2} \vect{x} - \frac{\sqrt{3}}{2} \vect{y}$, we obtain the
bandstructure $\phi (k_x,k_y)$ as:
\begin{multline}
e^{i 6\phi} e^{i 3 \varphi} + e^{i 3\phi} \cos ^3 \theta \left\{ \vphantom{\cos \left(\frac{3}{2} k_x \right)} e^{i 3 \xi}- e^{i 3 (\varphi-\xi)}  \right. \\  
\qquad \left. + \tan^2 \theta e^{i \varphi} \left[2 \cos \left(\frac{3k_x}{2} \right) \left( e^{i \xi} e^{i \frac{\sqrt{3}}{2} k_y} - e^{i(\varphi-\xi)}  e^{-i \frac{\sqrt{3}}{2} k_y} \right)\right. \right.
\left. \left. - e^{i(\varphi-\xi)} e^{i \sqrt{3} k_y} + e^{i \xi} e^{-i \sqrt{3} k_y} \vphantom{\cos \left(\frac{3k_x}{2} \right)} \right] \right\} -1 = 0.
\label{eq:bandstruct}
\end{multline}
Note that this expression does not depend on $\chi$.  Since
Eq.~(\ref{eq:bandstruct}) is a quadratic polynomial in $e^{i 3 \phi}$,
a bandgap closing at some point $(k_x^0,k_y^0)$ in the Brillouin zone
corresponds to a vanishing value of its discriminant, i.e.~(at least)
two roots being degenerate.  The locations of such bandgap closings in
the $(\theta,\xi,\varphi,\chi)$ parameter space of the system define
boundaries between different insulator phases, which may have
different topological order.  Fig.~\ref{fig:honphasediag} shows a
slice of the phase diagram of the honeycomb network model for
$\varphi=\chi=0$.

For $\varphi=0$ (which corresponds to $\mathrm{det}[S] = 1$), we can
simplify Eq.~\eqref{eq:bandstruct} to
\begin{equation}
e^{i 6\phi} + e^{i 3\phi} \cos ^3 \theta \left\{ \vphantom{\cos \left(\frac{3}{2} k_x \right)} 2 i \sin ( 3 \xi ) 
+ \tan ^2 \theta  \left[2 \cos \left(\frac{3k_x}{2} \right) 2 i \sin \left( \frac{\sqrt{3}}{2} k_y + \xi \right) - 2 i \sin ( \sqrt{3} k_y - \xi ) \vphantom{\cos \left(\frac{3}{2} k_x \right)} \right] \right\} -1 = 0,
\label{eq:bandstructphiz}
\end{equation}
and setting $\xi = \pi /2$ enables us to further simplify the
bandstructure equation to obtain:
\begin{equation}
e^{i 6\phi} + e^{i 3\phi} \cos ^3 \theta \left\{ \vphantom{\cos \left(\frac{3}{2} k_x \right)} - 2 i  + \tan ^2 \theta  \left[4 i \cos \left(\frac{3}{2} k_x \right) \cos \left( \frac{\sqrt{3}}{2} k_y \right)+ 2 i \cos ( \sqrt{3} k_y ) \vphantom{\cos \left(\frac{3}{2} k_x \right)} \right] \right\} -1 = 0.
\end{equation}
\end{widetext}
Defining $f(\vect{k})$ as
\begin{equation}
 f(\vect{k}) \equiv 4 \cos \left(\frac{3}{2} k_x \right) \cos \left( \frac{\sqrt{3}}{2} k_y \right) + 2 \cos ( \sqrt{3} k_y ),
\end{equation}
we obtain:
\begin{equation}
e^{i 6\phi} + e^{i 3\phi} \cos ^3 \theta i \left( \vphantom{\cos \left(\frac{3}{2} k_x \right)} - 2 + \tan ^2 \theta  f(\vect{k}) \vphantom{\cos \left(\frac{3}{2} k_x \right)}  \right) -1 = 0.
\end{equation}
In the tight-binding regime ($\theta \approx 0$), this gives
\begin{equation}
 e^{i 3 \phi_\pm} \approx \pm \theta \sqrt{ 3 + f(\vect{k})} + i,
\end{equation}
which yields
\begin{equation}
 \phi_\pm \propto \pm \frac{\theta}{3} \sqrt{3 + f(\vect{k})},
\end{equation}
in agreement with the standard result for the tight-binding
Hamiltonian of graphene when only the nearest-neighbor coupling is
taken into account \cite{castroneto}.  The coefficient $\theta/3$
plays the role of the nearest-neighbor hopping energy.

Using Eq.~(\ref{hexbands}), we can compute projected quasi-energy
bandstructures of the honeycomb network in the strip geometry, such as
those shown in Fig.~\ref{fig:hexbands} for zig-zag edges.  We have
also verified that similar edge states are present for armchair edges.


\begin{thebibliography}{99}
\bibitem{TKNN} D.~Thouless, M.~Kohmoto, M.~P.~Nightingale, and M.~den
  Nijs, Phys.~Rev.~Lett.~{\bf 49}, 405 (1982).
\bibitem{MStone} M.~Stone, \textit{Quantum Hall Effect} (World
  Scientific, 1992).
\bibitem{Moore} J.~E.~Moore, Nature {\bf 464}, 194 (2010).
\bibitem{RMPHasan} M.~Z.~Hasan and C.~L.~Kane, Rev.~Mod.~Phys. {\bf 82}, 3045 (2010).
\bibitem{RMPQi} X.-L.~Qi and S.-C.~Zhang, Rev.~Mod.~Phys. {\bf 83}, 1057 (2011).
\bibitem{Oka} T.~Oka and H.~Aoki, Phys.~Rev.~B {\bf 79}, 081406
  (2009).
\bibitem{Inoue} J.~Inoue and A.~Tanaka, Phys.~Rev.~Lett.~{\bf 105},
  017401 (2010).
\bibitem{Demler0} T.~Kitagawa, M.~S.~Rudner, E.~Berg, and E.~Demler,
  Phys.~Rev.~A {\bf 82}, 033429 (2010)
\bibitem{Demler} T.~Kitagawa, E.~Berg, M.~Rudner, and E.~Demler,
  Phys.~Rev.~B {\bf 82}, 235114 (2010).
\bibitem {Lindner} N.~H.~Lindner, G.~Refael and V.~Galitski, Nature Physics {\bf 7}, 490-495 (2011).
\bibitem{Gu} Z.~Gu, H.~A.~Fertig, D.~P.~Arovas, and A.~Auerbach, Phys.~Rev.~Lett.~{\bf 107}, 216601 (2011).
\bibitem{Cayssol} J.~Cayssol, B.~D\'ora, F.~Simon and R.~Moessner, Phys. Status Solidi RRL {\bf 7}, 101 (2013).
\bibitem{Levin} M.~S.~Rudner, N.~H.~Lindner, E.~Berg, and M.~Levin,
  Phys.~Rev.~X {\bf 3}, 031005 (2013).
\bibitem{Raghu1} F.~D.~M.~Haldane and S.~Raghu, Phys.~Rev.~Lett.~{\bf
  100}, 013904 (2008).
\bibitem{Raghu2} S.~Raghu and F.~D.~M.~Haldane, Phys.~Rev.~A {\bf 78},
  033834 (2008).
\bibitem{Wang1} Z.~Wang, Y.~D.~Chong, J.~D.~Joannopoulos, and
  M.~Solja\u{c}i\'{c}, Phys.~Rev.~Lett.~{\bf 100}, 013905 (2008).
\bibitem{Wang2} Z.~Wang, Y.~D.~Chong, J.~D.~Joannopoulos, and
  M.~Solja\u{c}i\'{c}, Nature {\bf 461}, 772 (2009).
\bibitem{LeHur}  J.~Koch, A.~A.~Houck, K.~Le~Hur, and S.~M.~Girvin, Phys.~Rev.~A {\bf 82}, 043811 (2010).
\bibitem{LeHur1} A.~Petrescu, A.~A.~Houck, and K.~Le~Hur, Phys.~Rev.~A {\bf 86}, 053804 (2012).
\bibitem{Khanikaev} A.~B.~Khanikaev, S.~H.~Mousavi, W.-K.~Tse,
  M.~Kargarian, A.~H.~MacDonald, and G.~Shvets, Nature Materials,
  doi:10.1038/nmat3520 (2012).
\bibitem{hafezi} M.~Hafezi. E.~A.~Demler, M.~D.~Lukin, and
  J.~M.~Taylor, Nature Phys.~{\bf 7}, 907 (2011).
\bibitem{hafezi2} M.~Hafezi. J.~Fan, A.~Migdall, and J.~M.~Taylor,
  arxiv:1302.2153.
\bibitem{Liang} G.~Q.~Liang and Y.~D.~Chong, Phys.~Rev.~Lett.~{\bf
  110}, 203904 (2013).
\bibitem{Fan} K.~Fang, Z.~Yu, and S.~Fan, Nature Phot.~{\bf 6},
  782 (2012).
\bibitem{Szameit} M.~C.~Rechtsman, J.~M.~Zeuner, Y.~Plotnik, Y.~Lumer,
  S.~Nolte, M.~Segev, and A.~Szameit, Nature {\bf 496}, 196 (2013).
\bibitem{ChalkerCo} J.~T.~Chalker, and P.~D.~Coddington, J.~Phys.~C {\bf 21}, 2665 (1988).
\bibitem{Kivelson} D.-H.~Lee, Z.~Wang, and S.~A.~Kivelson,
  Phys.~Rev.~Lett. {\bf 70}, 4130 (1993).
\bibitem{Lee} D.-H.~Lee, Phys.~Rev.~B {\bf 50}, 10788 (1994).
\bibitem {Kramer} B.~Kramer, T.~Ohtsuki, and S.~Kettemann, Phys.~Rep. {\bf 417}, 211 (2005).
\bibitem{klessemet1} R.~Klesse, and M.~Metzler, Europhys. Lett. {\bf 32}, 229 (1995).
\bibitem{HoChalker} C.-M.~Ho, and J.~T.~Chalker, Phys.~Rev.~B {\bf 54}, 8708 (1996).
\bibitem{klessemet2} R.~Klesse, and M.~Metzler, Int. J. Mod. Phys. C {\bf 10}, 577 (1999).
\bibitem{janssen} M.~Janssen, M.~Metzler, and M.~R.~Zirnbauer, Phys. Rev. B {\bf 59}, 15836 (1999).
\bibitem{janssenbook} M.~Janssen, \textit{Fluctuations and Localization In Mesoscopic Electron Systems} (World Scientific, 2001).
\bibitem{Laughlin} R.~B.~Laughlin, Phys.~Rev.~B {\bf 23}, 5632 (1981).
\bibitem{Brouwer} D.~Meidan, T.~Micklitz, and P.~W.~Brouwer,
  Phys.~Rev.~B {\bf 84}, 195410 (2011).  See also I.~C.~Fulga, F.~Hassler, and A.~R.~Akhmerov, Phys.~Rev.~B {\bf 85}, 165409 (2012).
\bibitem{haldane88} F.~D.~M.~Haldane, Phys. Rev. Lett. {\bf 61}, 2015 (1988).
\bibitem{yariv02} A.~Yariv, IEEE Phot.~Tech.~Lett.~{\bf 14}, 483
  (2002).
\bibitem{yariv_wg} J.~K.~S.~Poon, J.~Scheuer, S.~Mookherjea,
  G.~T.~Paloczi, Y.~Huang, and A.~Yariv, Opt.~Ex.~{\bf 12}, 90 (2004).
\bibitem{Schnyder08} A.~P.~Schnyder, S.~Ryu, A.~Furusaki, and A.~W.~W.~Ludwig, Phys.~Rev.~B {\bf 78}, 195125 (2008).
\bibitem{Schnyder09} A.~P.~Schnyder, S.~Ryu, A.~Furusaki, and A.~W.~W.~Ludwig, AIP~Conf.~Proc.~{\bf 1134}, 10 (2009).
\bibitem{Kitaev} A.~Kitaev, AIP~Conf.~Proc.~{\bf 1134}, 22 (2009).
\bibitem{ChalkerDoh} J.~T.~Chalker, and A.~Dohmen, Phys.~Rev.~Lett. {\bf 75}, 4496 (1995).
\bibitem{obuse1} H.~Obuse, A.~Furusaki, S.~Ryu, and C.~Mudry, Phys. Rev. B {\bf 76}, 075301 (2007).
\bibitem{obuse2} H.~Obuse, A.~Furusaki, S.~Ryu, and C.~Mudry, Phys. Rev. B {\bf 78}, 115301 (2008).
\bibitem{ryu} S.~Ryu, C.~Mudry, H.~Obuse, and A.~Furusaki, New J. Phys. {\bf 12}, 065005 (2010).
\bibitem{castroneto} A.~Castro Neto, F.~Guinea, N.~M.~R.~Peres, K.~S.~Novoselov, and A.~K.~Geim, Rev. Mod. Phys. {\bf 81}, 109 (2009).
\bibitem{KitagawaNatCom} T.~Kitagawa, M.~A.~Broome, A.~Fedrizzi, M.~S.~Rudner, E.~Berg, I.~Kassal, A.~Aspuru-Guzik, E.~Demler, and A.~G.~White, Nature Commun. {\bf 3}, 882 (2012).
\bibitem{Segev} T.~Schwartz, G.~Bartal, S.~Fishman, and M.~Segev, Nature (London) {\bf 446}, 52 (2007).
\bibitem{Segevreview} M.~Segev, Y.~Silberberg, and D.~N.~Christodoulides, Nat. Photonics {\bf 7}, 197 (2013).
\bibitem{localizationRMP} F.~Evers and A.~D.~Mirlin, Rev.~Mod.~Phys.~{\bf 80}, 1355 (2008).
\end{thebibliography}
\end{document}